\documentclass[a4paper]{jpconf}
\usepackage{graphicx}

\def\beq{\begin{equation}}
\def\eeq{\end{equation}}
\def\be{\begin{eqnarray}}
\def\ee{\end{eqnarray}}

\begin{document}
\title{Confronting electron- and neutrino-nucleus scattering}

\author{Omar Benhar}

\address{INFN and Department of Physics \\ 
``Sapienza'' Universit\`a di Roma. I-00185 Roma, Italy}

\ead{omar.benhar@roma1.infn.it}

\begin{abstract}
The analysis of the sample of charged current
quasi elastic events collected by the MiniBooNE Collaboration suggests 
that the scheme successfully employed to describe
electron-nucleus scattering fails to explain neutrino-nucleus cross sections. 
I argue that, due to flux average, the double differential neutrino-nucleus cross 
section does not allow for a clearcut determination of the dominant reaction mechanism.
A systematic study of the large body of electron scattering data
may help to identify  the processes, other than single nucleon knockout,
contributing to the observed neutrino cross section.
\end{abstract}

\begin{figure}[h]
\begin{minipage}{18pc}
\includegraphics[width=18pc]{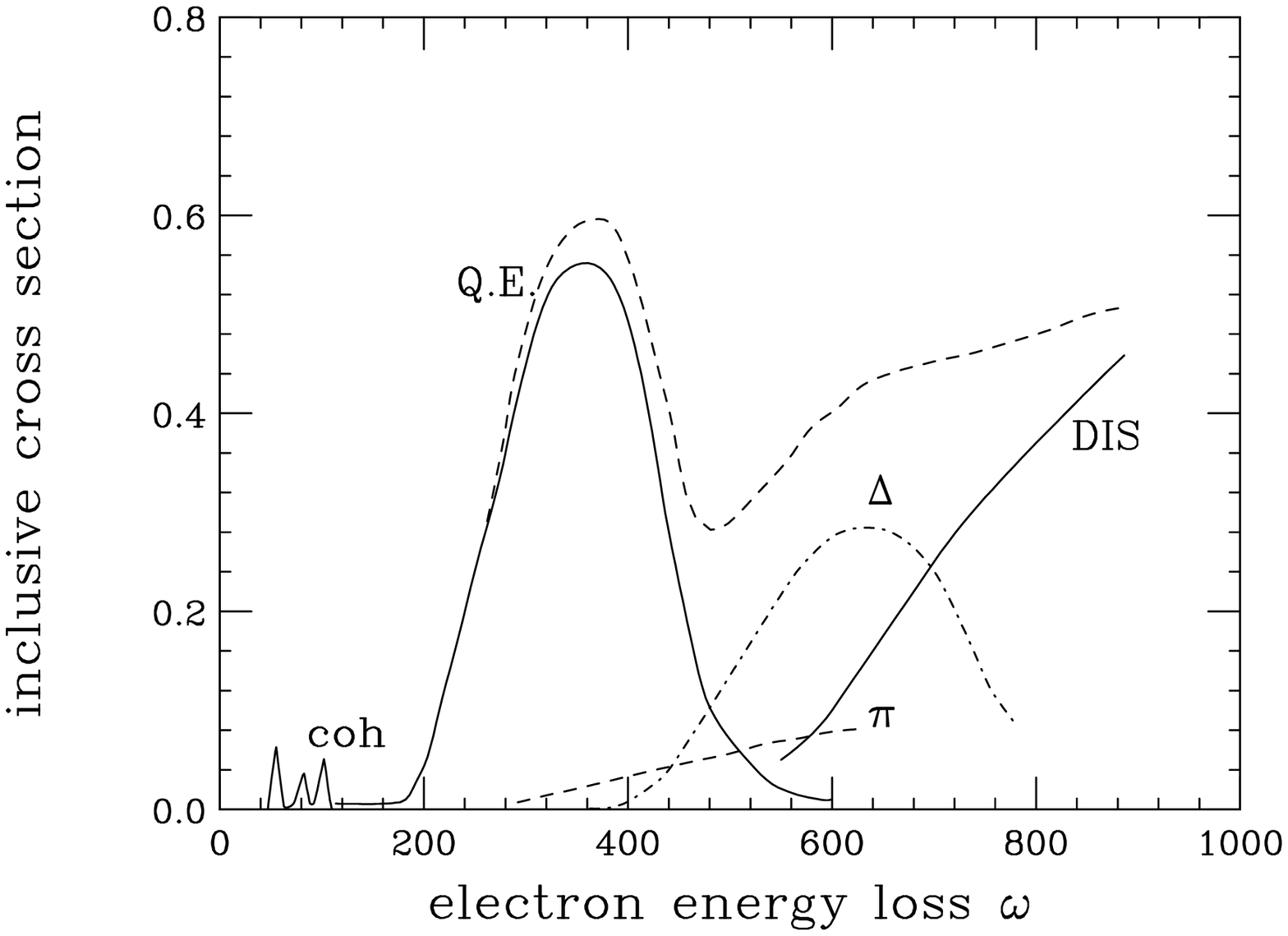}
\caption{\label{xsec_ee_1}Typical behavior of the inclusive electron-nucleus scattering cross section at beam
 energy around 1 GeV, as a function of the electron energy loss $\omega$.}
\end{minipage}\hspace{2pc}%
\begin{minipage}{18pc}
\includegraphics[width=17pc]{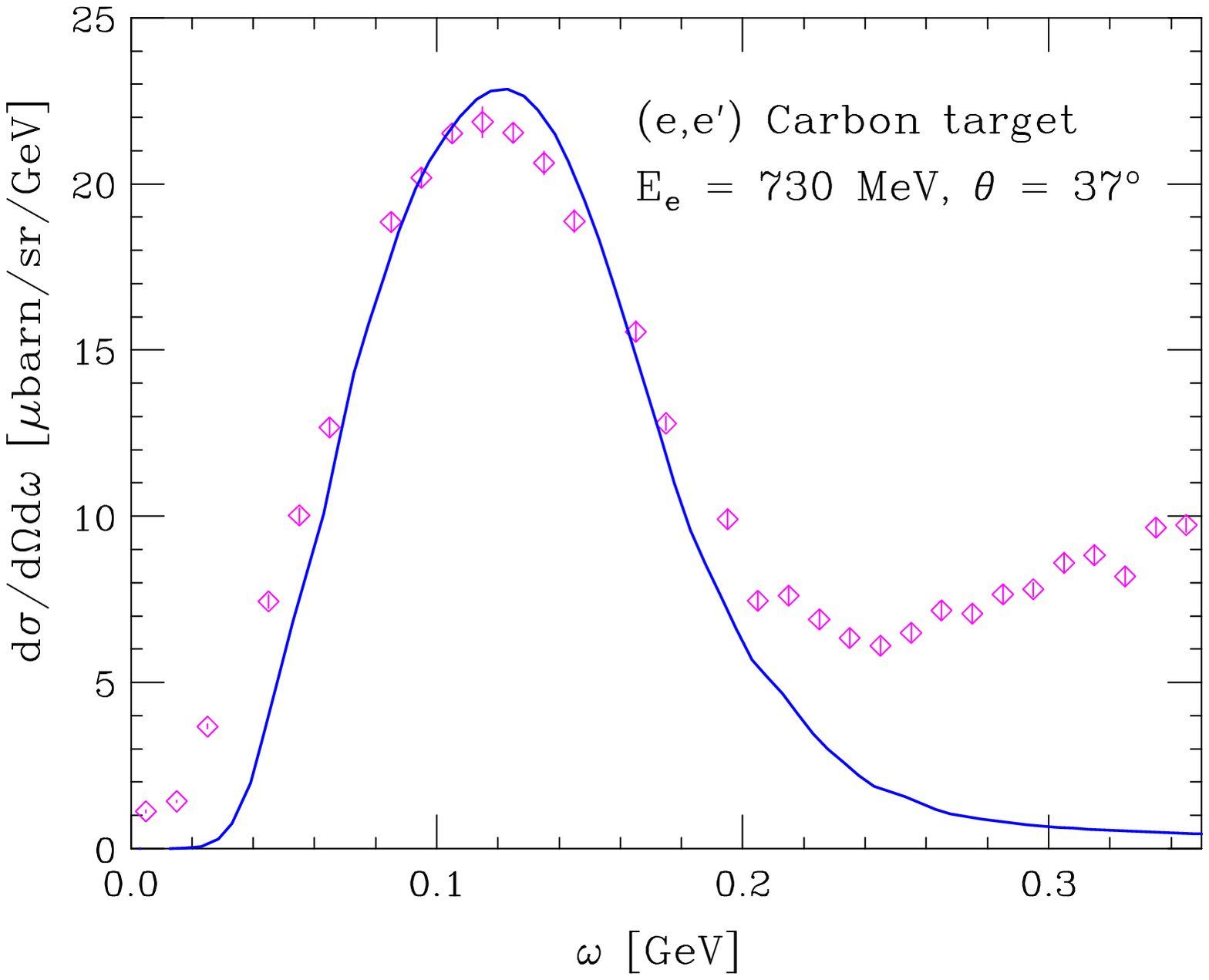}
\caption{\label{xsec_ee_2}Quasi elastic electron-carbon cross section at beam energy $E_e=$ 730 MeV and electron scattering
angle $\theta_e=37^\circ$, plotted as a function of the energy loss $\omega$ \cite{PL}. The data points are from
Ref. \cite{12C_ee}.}
\end{minipage}
\end{figure}

Electron-nucleus scattering cross sections are usually analyzed at fixed beam energy, $E_e$, and
electron scattering angle, $\theta_e$, as a function of the electron energy loss $\omega$.
As an example, Fig.~\ref{xsec_ee_1} shows the typical behavior of the inclusive cross sections at beam
energy around 1 GeV. It is apparent that the different reaction mechanisms, yielding the dominant contributions
to the cross section at different values of $\omega$ (corresponding to different values of the
Bjorken scaling variable $x=Q^2/2m\omega$, where $m$ is the nucleon mass and
$Q^2=4E_e(E_e-\omega)\sin^2 \theta_e/2$) can be easily identified.

The bump centered at $\omega \sim Q^2/2m$, or $x \sim 1$, the position and width of which are determined by the
momentum and removal energy distribution of the struck particle,  corresponds to single nucleon
knockout, while the structure visible at larger $\omega$, or lower $x$, reflects the onset of coupling to two-nucleon
currents, arising from meson exchange processes, excitation of nucleon resonances and
deep inelastic scattering.

The quasi elastic electron-nucleus cross section can be described at quantitative level within the
Impulse Approximation (IA), applicable when the magnitude of the three-momentum transfer,  $|{\bf q}|$,
is small compared to the average distance between nucleons in the target nucleus \cite{RMP}.
As an example, Fig. \ref{xsec_ee_2} shows the inclusive electron-carbon cross section at beam energy $E_e=$ 730~MeV
and electron scattering angle $\theta_e=37^\circ$, plotted as a function of the energy loss $\omega$ \cite{PL}.
The data points are taken from Ref. \cite{12C_ee}, while the theoretical results have been obtained within
the approach described in Refs. \cite{RMP,PRD}, using a state-of-the-art parametrization of the vector nucleon form factors, 
extracted from the measured electron-proton electron-deuteron cross sections (for a review, see, e.g., Ref.~\cite{VFF}).

\begin{figure}[h]
\begin{minipage}{18pc}
\includegraphics[width=16pc]{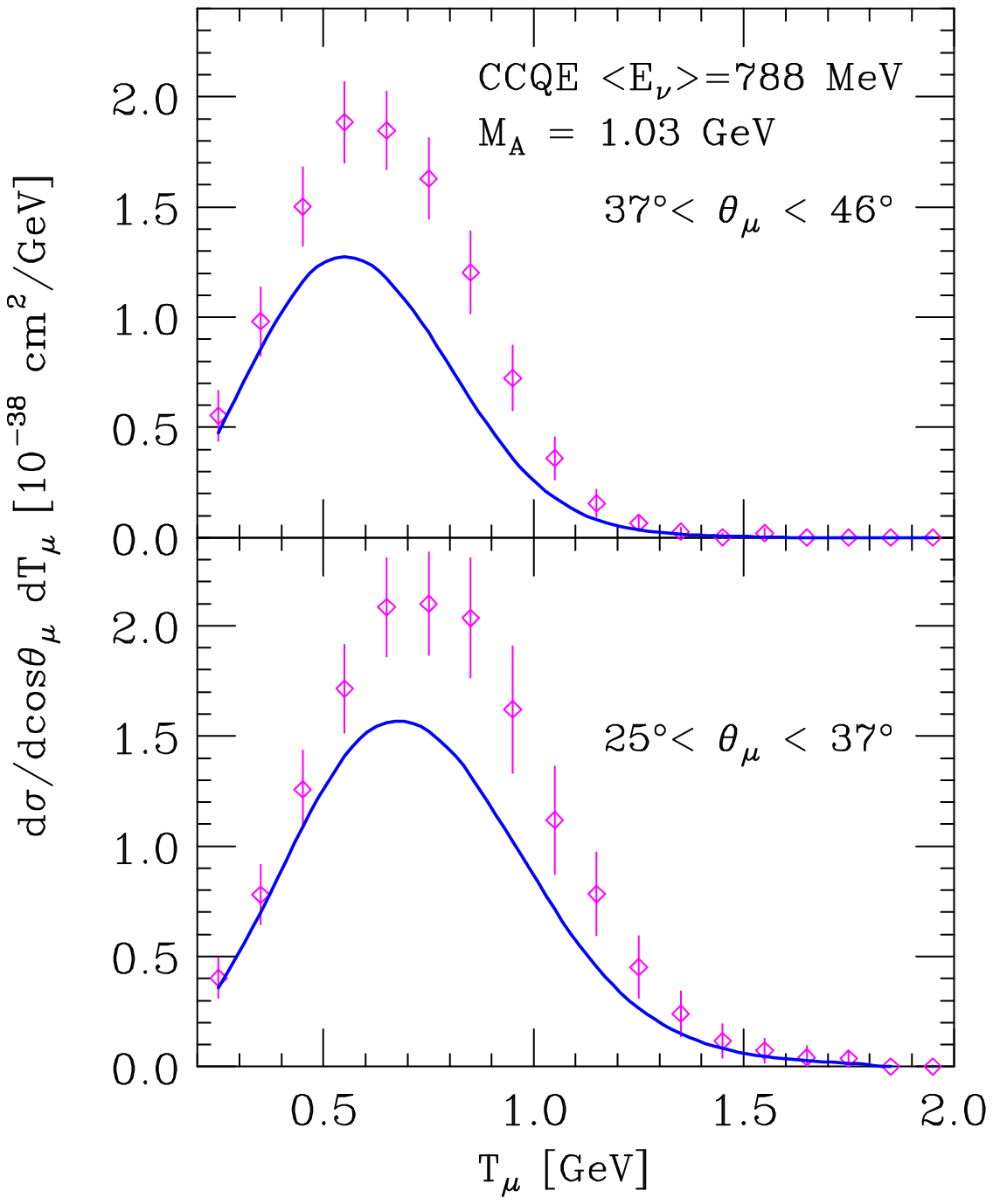}
\caption{ \label{dsigma}Flux averaged double differential CCQE cross section measured by the MiniBooNE collaboration
\cite{BooNECCQE},  shown as a function of the kinetic energy of the outgoing muon. The upper and lower panels correspond to
different values of the muon scattering angle. The theoretical results have been obtained in Ref.\cite{PL}, using the formalism described in
Refs. \cite{RMP,PRD}, the same vector form factors employed in the calculation of the electron scattering cross section of Fig. \ref{xsec_ee_2},
and a dipole parametrizaition of the axial form factor with $M_A=1.03$ MeV.}
\end{minipage}\hspace{2pc}%
\begin{minipage}{18pc}
\includegraphics[width=18pc]{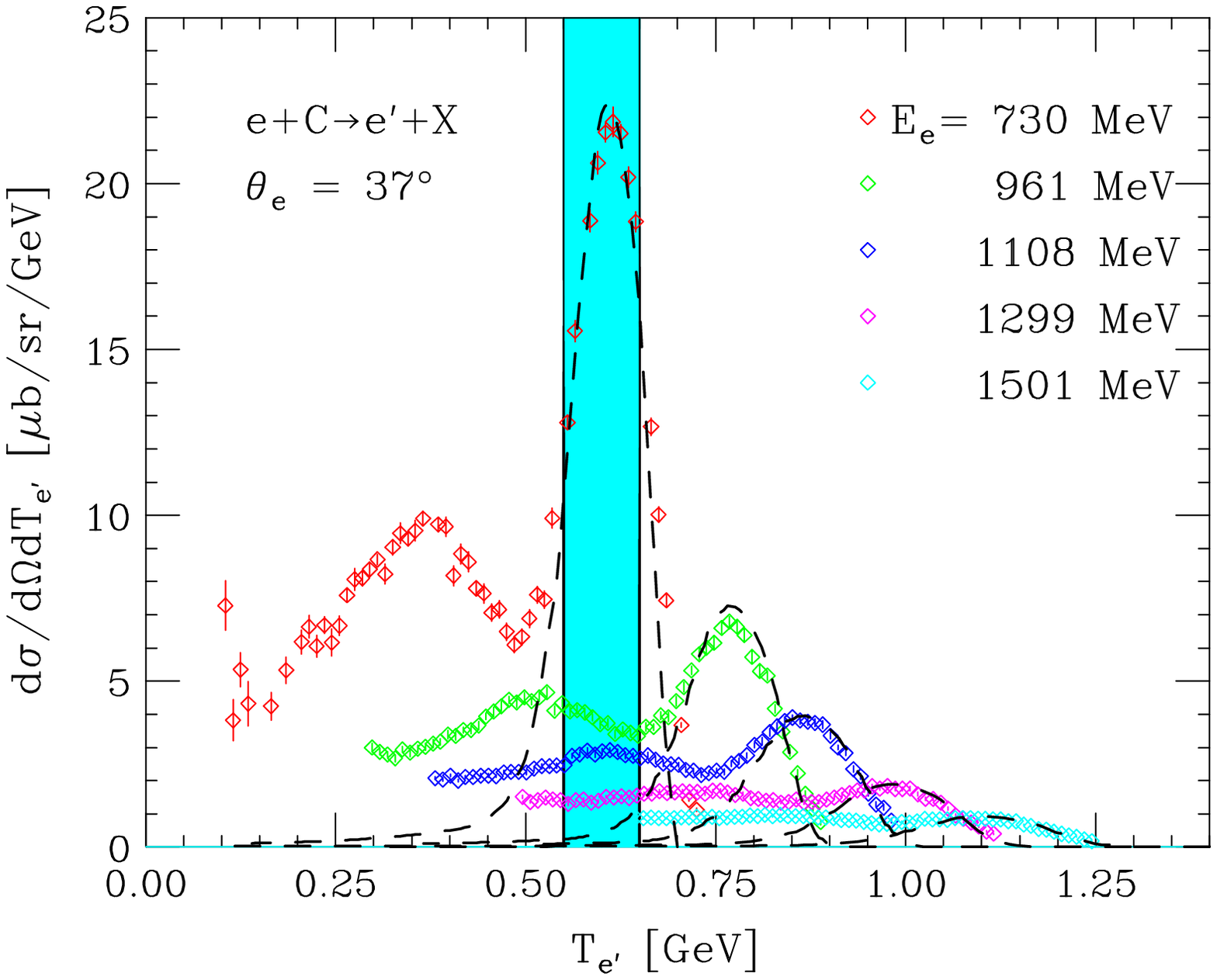}
\caption{\label{newdist} Inclusive electron-carbon cross sections at $\theta_e=$ 37 deg and beam energies ranging
 between 0.730 and 1.501 GeV \cite{12C_ee,12C_ee_2}. The dashed lines represent the single nucleon knock out contribution,
 computed within the approach described in Refs. \cite{RMP,PRD}. The shaded area shows the
  energy bin corresponding to the peak of the cross section of Fig. \ref{xsec_ee_2}.}
\end{minipage}
\end{figure}

Applying the same scheme employed to obtain the solid line of Fig. \ref{xsec_ee_2} to neutrino scattering
one gets the results shown in Fig. \ref{dsigma}. The data points represent the double differential CCQE cross section of 
 Ref.~\cite{BooNECCQE} averaged over the MiniBooNE neutrino flux, the mean energy of which is \
$\langle~E_\nu~\rangle~=~788$~MeV, 
plotted as a function of the kinetic energy of the outgoing muon at different values of the muon scattering angle. The solid lines show
the results of calculations performed using the same nuclear spectral
functions and  vector form factors employed in the calculation of the electron scattering cross section of Fig.~\ref{xsec_ee_2} 
and the dipole parametrization of the axial form factor, with the axial mass $M_A=1.03$~MeV \cite{PL} obtained from the world average 
of low statistics deuterium data \cite{bernard,bodek,nomad}.

Comparison of Figs. \ref{xsec_ee_2} and \ref{dsigma} indicates that the electron and neutrino cross sections corresponding to the same target and
{\em seemingly} comparable kinematical conditions (the position of the QE peak in Fig.~\ref{xsec_ee_2} corresponds to
kinetic energy of the scattered electron $\sim~610$~MeV) cannot be explained using the same theoretical approach
and the value of the axial mass resulting from deuterium measurements.

The authors of Refs.~\cite{BooNECCQE,K2K,BOONE} suggested that the measured CCQE cross sections 
may be explained by advocating a larger value of $M_A$,  that should be regarded as an {\em effective} axial mass, modified by nuclear effects not included in the
Fermi gas model employed in data analysis. However, most theoretical models of the CCQE cross section
(for recent reviews see Ref.\cite{nuint09}) fail to support this explanation. In addition, the analysis of the large body of electron scattering data
provide overwhelming evidence that the vector form factors are {\em not} modified in the nuclear medium.

In Refs.~\cite{PL,neutrino_2010} it is argued that the interpretation of the data sample of Ref.~\cite{BooNECCQE} as CCQE events may
be hampered by the fact that,  as the energy of the incoming beam is not fixed, 
the observed energy of the outgoing charged lepton {\em does not} uniquely determine the energy transfer to the target, 

The implications of flux average can be easily understood considering  the neutrino cross section
at muon energy corresponding to the maximum of the spectrum shown in the upper panel of
Fig. \ref{dsigma}, i.e. $T_\mu = 0.55$ GeV and $\cos \theta$ = 0.75. In this kinematics,
$x=1$ and 0.5 correspond to neutrino energy $E_\nu=$ 0.788 and 0.975 GeV, respectively.
As the values of the MiniBooNE flux corresponding to these energies are within less
than 20\% of one another, flux integration leads to collect contributions from different
regimes, i.e. different reaction mechanisms, with about the same probability.

This feature can be best illustrated using the measured electron-carbon scattering cross sections.
Figure  \ref{newdist} shows the data from Refs. \cite{12C_ee,12C_ee_2}, taken at electron
scattering angle $\theta_e=$~37~deg and beam energies ranging between 0.730 and 1.501 GeV, plotted
as a function of the energy of the outgoing electron. It clearly appears that the energy bin corresponding to
the top of the quasi elastic peak at $E_e = 0.730$ GeV, shown by the shaded area, receives significant contributions from
 cross sections corresponding to different beam energies and different values of $x$.

The data displayed in Fig.~\ref{newdist} strongly suggest that the description of the flux-integrated CCQE neutrino nucleus
cross section requires the inclusion of reaction mechanisms other than single nucleon knockout.
According to the authors of Ref. \cite{martini,nieves}, the most important competing mechanism is
multinucleon knockout, leading to two particle-two hole (2p2h) final states.  Note that in neutrino
experiments this final states cannot be distinguished from the one particle-one hole final states
associated with single nucleon knockout.

Multinucleon knockout is known to occur due to  i) initial state nucleon-nucleon correlations, ii) final state interactions between the struck 
nucleon and the spectator particles and iii) coupling to the two-body
nuclear electroweak current. 

Correlations between nucleons in the target ground state give rise to the tail extending to large $\omega$,
clearly visible in Fig. \ref{xsec_ee_2}. However, their contribution, strongly constrained by semi-inclusive $(e,e^\prime p)$ data \cite{daniela}
 turns out to be quite small (less than 10\% of the integrated spectrum). In principle, this reaction mechanism might be clearly identified detecting 
 two nucleons moving in opposite directions with momenta much larger than the Fermi momentum ($\sim$ 250 MeV). 

Final state interactions are not expected to play a relevant relevant in this context, as their main effects, which amounts to
a shift and a redistribution of the inclusive strength, mostly affects the region of low energy loss, corresponding to
$x>1$.

The most important correction is likely to arise from processes involving the nuclear two-body current, as advocated by the authors of
Refs. \cite{martini,nieves}. It is long known that inclusion of these processes is needed to explain the nuclear electromagnetic
response in the transverse channel \cite{joe}.  In addition, their contribution turns out to decrease as the momentum
transfer increases. This behavior may explain why the CCQE data at high neutrino energies collected by the NOMAD
collaboration \cite{nomad} can be described without any modification of the value of the nucleon axial mass.

In conclusion, comparison between electron and neutrino scattering data suggest that the analysis of the flux-averaged neutrino 
cross sections requires the development of models including a variety of relevant reaction 
mechanisms. While the approaches developed in Refs.\cite{martini,nieves} certainly represent an important step towards the achievement 
of this goal,  more theoretical work is still needed, e.g. to extend the applicability of the models to the regions of pion production and 
deep inelastic scattering. 

The critical requirement to be met in developing new theoretical approaches will be consistency, i.e. the ability to describe different 
kinematical regimes using the same dynamical model. Within {\em ab initio} nuclear many-body theory this amounts to requiring that the 
target initial and final states be obtained from the same hamiltonian, fitted to the properties of exactly solvable few-nucleon systems, 
which also largely determines the structure of the vector two-body current through the continuity equation. 

\subsection*{Acknowledgments}
The results discussed in this paper have been obtained in collaboration with Pietro Coletti and
Davide Meloni. An illuminating discussion with Georgia Karagiorgi and Camillo Mariani is also
gratefully acknowledged. Finally, I would like  to express to the organizers of NUFACT11 my
deepest appreciation for setting up a most enjoyable and fruitful meeting.


\section*{References}
\begin{thebibliography}{99}

\bibitem{RMP}
Benhar O, Day D and Sick I  2008 \emph{Rev. Mod. Phys.}  \textbf{80} 189

\bibitem{PL}
Benhar O, Coletti P and Meloni D 2010  \emph{Phys. Rev. Lett.} \textbf{105} 132301

\bibitem{12C_ee}
O$^\prime$Connell J S \textit{et al.} 1987  \emph{Phys Rev. C} \textbf{35} 1063

\bibitem{PRD}
Benhar O, Farina N, Nakamura H, Sakuda M and Seki R 2005 \emph{Phys Rev. D}
  \textbf{72} 053005
  
 \bibitem{VFF}
Perdrisat C, Punjabi V and Vanderhaeghen M 2007  \emph{Prog. Part. Nucl. Phys.}
  \textbf{59} 694

\bibitem{BooNECCQE}
Aguilar-Arevalo A A \textit{ et al.} (MiniBooNE Collaboration) 2010 \emph{Phys.
  Rev. D} \textbf{81} 092005

\bibitem{bernard}
Bernard  V \textit{et~al.} 2002 \emph{Phys. Phys. G} \textbf{28} R1 

\bibitem{bodek}
Bodek A, Avvakumov S , Bradford R and Budd H, 2008 \emph{Eur. Phys .J. C}
  \textbf{53} 349

\bibitem{nomad}
Lyubushkin V \textit{et~al.} (NOMAD~Collaboration) (2009) \emph{Eur. Phys .J. C} \textbf{63} 355

\bibitem{K2K}
Gran R \textit{et~al.} (K2K~Collaboration) 2006 \emph{Phys. Rev. D} \textbf{74} 052002

\bibitem{BOONE}
Aguilar-Arevalo A A \textit{ et al.} (MiniBooNE Collaboration) 2008 \emph{Phys. Rev. D} \textbf{100} 
032301 

\bibitem{nuint09}
Sanchez F, Sorel M and Alvarez-Ruso M (ed) 2010 \emph{Proceedings of the
  Sixth International Workshop on Neutrino-Nucleus Interactions in the Few-GeV
  Region (NUINT-09)}, AIP Conf. Proc. {\bf 1189}
  
\bibitem{neutrino_2010}
Benhar O 2010 arXiv:1012.2032v1 [nucl-th]    

\bibitem{12C_ee_2}
Sealock R M \textit{ et~al.} 1989 \emph{Phys Rev. Lett.} \textbf{62} 1350  

\bibitem{martini}
Martini M, Ericson M, Chanfray G and Marteau J 2010 \emph{Phys Rev. C} \textbf{81}
  045502 

\bibitem{nieves}
Nieves J  Ruiz-Simo I R and Vicente-Vacas M J 2011 arXiv:1106.5374v1 [hep-ph]

\bibitem{daniela}
Rohe D \textit{et al.} (JLAB E97-006 Collaboration) 2004 \emph{Phys. Rev. Lett.}
\textbf{93} 182501 

\bibitem{joe}
Carlson J and Schiavilla R 1998 \emph{Rev. Mod. Phys.} \textbf{70} 743 
  

\end{thebibliography}
\end{document}